\pdfoutput=1
\documentclass{PoS}

\usepackage{graphicx}
\usepackage{dcolumn}
\usepackage{bm}
\usepackage{psfrag}
\usepackage{xspace}
\usepackage{multirow}
\usepackage{lscape}
\usepackage{comment}
\usepackage{here}

\newcommand{\dzero}     {D\O}
\newcommand{\ttbar}     {\mbox{$t\bar{t}$}}
\newcommand{\ppbar}     {\mbox{$p\bar{p}$}}
\newcommand{\qqbar}     {\mbox{$q\bar{q}$}}
\newcommand{\ssbar}     {\mbox{$s\bar{s}$}}
\newcommand{\eplus}     {$e$+jets}
\newcommand{\muplus}     {$\mu$+jets}

\newcommand{\ljets}     {$\ell$+jets}

\newcommand{\fbmone}      {fb$^{-1}$}
\newcommand{\pbmone}      {pb$^{-1}$}

\newcommand{\ltau}      {\ensuremath{\tau \ell}}
\newcommand{\etau}      {\ensuremath{\tau e}}
\newcommand{\mutau}     {\ensuremath{\tau \mu}}

\newcommand{\ttb}{\mbox{$t\bar{t}$}}

\title{Top Quark Highlights}

\ShortTitle{Top Quark Highlights}

\author{\speaker{Christian Schwanenberger}\\
        University of Manchester\\
        E-mail: \email{schwanen@fnal.gov}}

\abstract{Highlights of top quark physics presented at the 2009 Europhysics Conference
on High Energy Physics from 16-22 July 2009 in Krakow, Poland, are
reviewed.}

\FullConference{The 2009 Europhysics Conference on High Energy Physics,\\
		 July 16 - 22 2009\\
		 Krakow, Poland}

\begin{document}

\section{Introduction}

At hadron colliders, top quarks can be produced in pairs via the
strong interaction or singly via the electroweak
interaction~\cite{singletop-willenbrock}. Top quarks were first
observed via pair production at the Fermilab Tevatron Collider in
1995~\cite{top-obs-1995}. Since then, pair production has been used to
make precise measurements of several top quark properties, including
the top quark mass~\cite{top-mass-properties}.

This review gives an overview of current measurements of top quark
cross sections, top quark properties and searches for new physics in
the top sector at the Tevatron proton-antiproton collider at a
center-of-mass energy of $\sqrt{s}=1.96$~TeV at Fermilab with datasets
corresponding to an integrated luminosity of up to 3.6~\fbmone. 
It also presents one result from the HERA $e^\pm p$ collider at a
center-of-mass energy of $\sqrt{s}=319$~GeV using a dataset of
474~\pbmone. 
Prospects for
top quark analyses in proton-proton scattering at the Large Hadron
Collider (LHC) are discussed as well.

\section{Top Quark Physics at the Tevatron}

The Tevatron is still the only place where top quarks can be produced
and studied directly. At the Tevatron, top  
quarks are either produced in pairs via the strong interaction or
singly via the electroweak interaction.
In the
framework of the Standard Model (SM), each top quark is expected to
decay nearly 100\% of the times into a $W$~boson 
and a $b$~quark~\cite{top-mass-properties}.
$W$ bosons can decay hadronically into
$q\bar{q}^\prime$ pairs or
leptonically into $e\nu_{e}$, $\mu\nu_{\mu}$ and $\tau\nu_{\tau}$   with
              the $\tau$ in turn decaying into an electron, a muon, or  hadrons,
              and associated neutrinos.

In \ttbar\ production, if both $W$ bosons decay hadronically the final state is called
all-hadronic (or all-jets) channel. If one of the $W$ bosons decays hadronically
while the other one produces 
a direct electron or muon  or a secondary electron or muon from
$\tau$ decay, the final state is 
referred to as the $\ell$+jets channel.
If both $W$ bosons  decay leptonically, this leads to a dilepton
($\ell\ell$) final
state containing a pair of electrons, a pair of muons, or 
an electron and a muon, or a hadronically decaying
tau accompanied either by an electron or a muon (the \ltau\ channel).

\subsection{Top Production Cross Sections}

Exploring the top cross section in different decay channels and using
different assumptions is important because signs of new
physics might appear differently in the various
channels. 

\subsubsection{Top Quark Pair Production}

The top quark pair production cross section $\sigma_{t\bar{t}}$ is
known to high accuracy in the Standard Model
(SM)~\cite{SMtheory,SMtheory_K,SMtheory_C}. The measurement of the top
quark pair production cross section therefore provides an important
test of such higher order QCD
calculations including soft gluon resummations. Any deviation from
the SM prediction of the measured \ttbar\ cross section could either
be a hint for new physics in top pair production or in top quark decays.
For example, an exotic decay of
the top quark, such as the decay into a charged Higgs boson and a $b$
quark ($t\rightarrow H^{+}b$) would lead to deviations
of the measured $\sigma_{t\bar{t}}$ in
individual final states compared to the SM prediction. 

There are two different techniques to enhance the top pair signal over
the background. In one method topological variables are used, in the
other method the identification of jets originated from $b$ quarks
($b$-tagging) is utilized. 

For the $\ell$+jets channel, the CDF collaboration developed a neural
network technique to maximize the discriminating power from kinematic
and topological variables using 2.8~\fbmone\ of data~\cite{cdf_xsec_nn}. A fit to a
neural network output constructed of those variables was performed.
The sensitivity of the neural network
technique is comparable to that for the traditional CDF secondary
vertex $b$-tag method~\cite{cdf_xsec_btag}, which suppresses the
dominant background 
from $W$+jets at a cost of a 45\% loss in signal efficiency.

The \ttbar\ cross section measurement using the topological method is
dominated by systematic uncertainties. The largest source is the uncertainty on the
luminosity determination which is 5.8\%. In order to significantly
reduce this uncertainty, the correlation between the
luminosity measurements in two different processes can be
exploited. Here the ratio of the \ttbar\ and $Z$ cross sections is
used where the luminosity uncertainty almost
entirely cancels out. Multiplying this ratio by the best theoretical calculation
of the $Z$ cross section, a \ttbar\ cross section can be obtained. In effect,
the luminosity uncertainty is replaced by the significantly smaller theoretical and
experimental uncertainties on the $Z$ cross section.
The measured \ttbar\ cross section is
$\sigma_{\ttbar} = 6.9 \pm 0.4 {\rm (stat)} \pm 0.4 {\rm (sys)} \pm
0.1 {\rm (theory)}$~pb for a top quark mass of 175~GeV. The total
uncertainty is 8\%. 
 
In the all-hadronic channel with 2.9~\fbmone\ of data collected with a
multijet trigger, for example,
$b$-tagging is used for the measurement of the \ttbar\ cross
section~\cite{cdf_allhad}. 
Using a dedicated neural network based kinematic selection, events
with one or at least 2 $b$-tags are studied using templates for the
reconstructed top quark mass simultaneously with an in situ  
measurement of the Jet Energy Scale (JES). 
The measured \ttbar\ production cross section amounts to 
$\sigma_{\ttbar} = 7.2 \pm 0.5 {\rm (stat)}  \pm 1.1 {\rm (syst)}  \pm
0.4 {\rm (lumi)}$~pb for a top mass of 172.5~GeV.

\begin{figure*}[t]
\centering
\includegraphics[width=70mm]{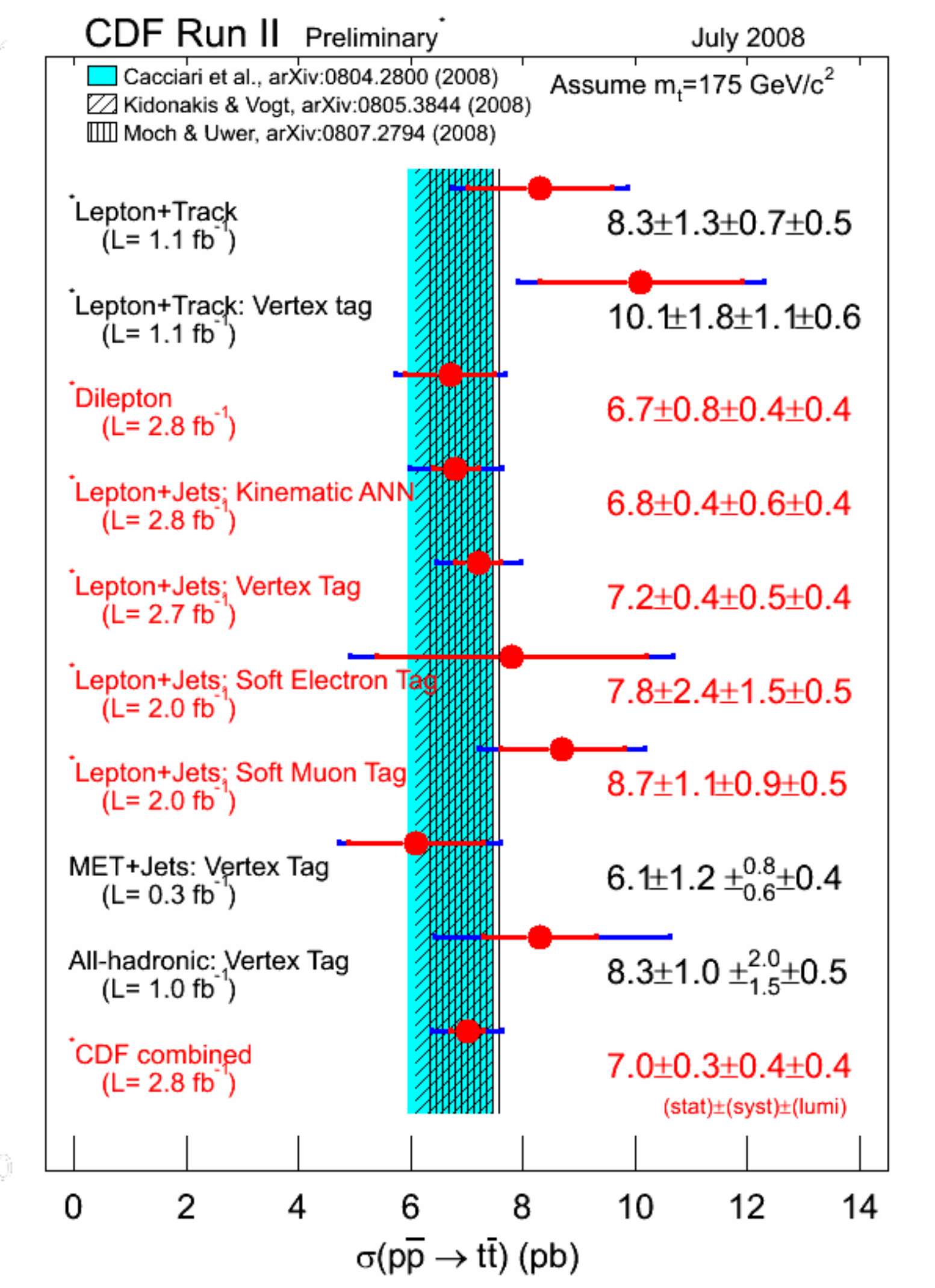}
\includegraphics[width=70mm]{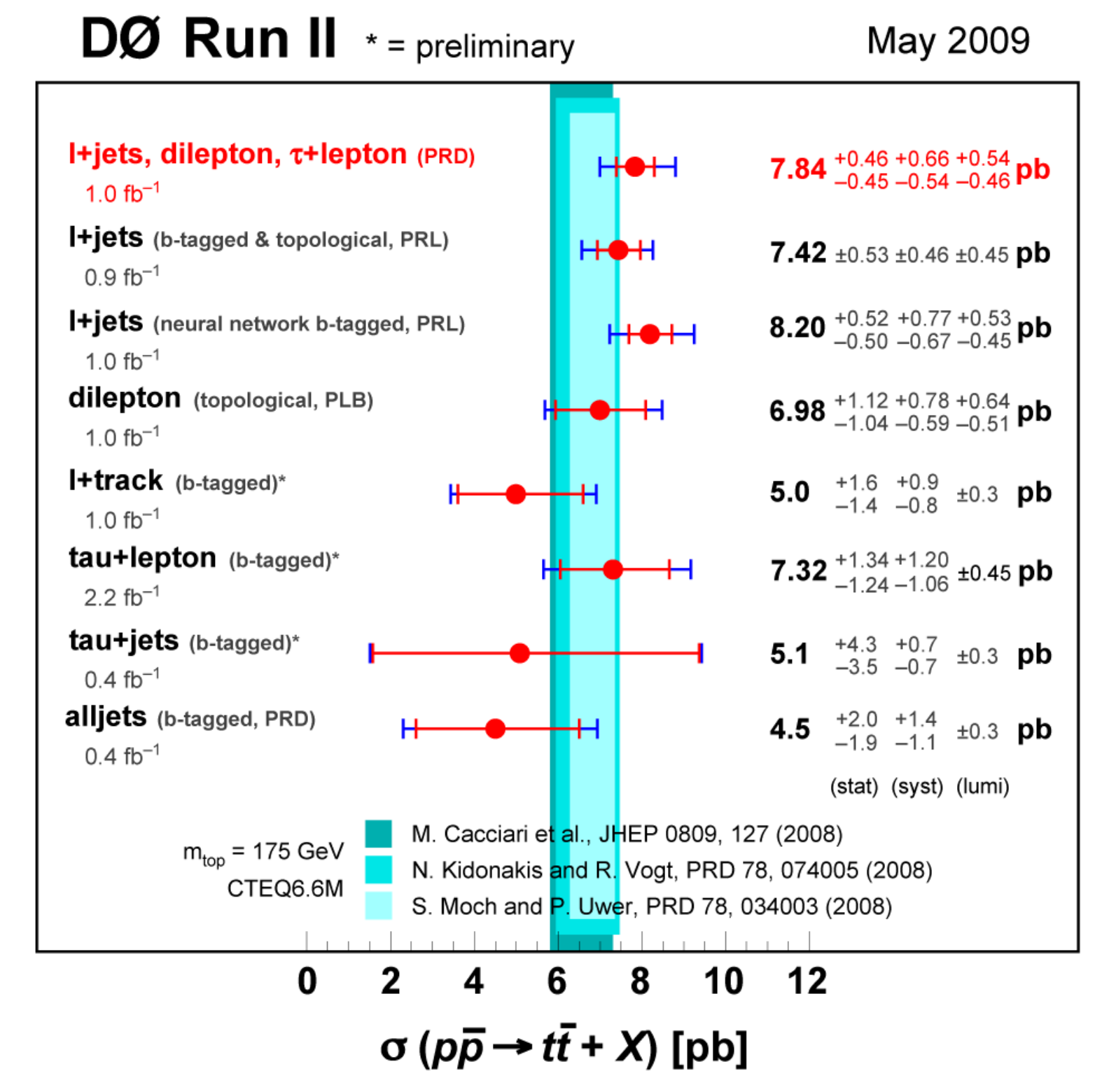}
\caption{Summary of \ttbar\ cross section measurements in different
  decay channels by the CDF (left) and the D\O\ Collaborations
  (right). The combination for each experiment is
displayed, too.
} \label{fig:xsec}
\end{figure*}

Figure~\ref{fig:xsec} summarizes the measurements of the \ttbar\ cross
section in different decay channels performed by the CDF
and D\O\ Collaborations. All measurements agree with each other and
agree with the SM predictions. 

\subsubsection{Single Top Quark Production}

Single top quark
production serves as a probe of the $Wtb$
interaction~\cite{singletop-wtb}, and its production cross section
provides a direct measurement of the magnitude of the quark mixing
matrix element $V_{tb}$ without assuming three quark
generations~\cite{singletop-vtb-jikia}. However, measuring the yield
of single top quarks is difficult because of the small production rate
and large backgrounds. At the Tevatron single top quarks are produced
by either a $t$-channel exchange of a virtual $W$ boson which combines with
a highly energetic $b$ quark to produce a top quark, or by an
$s$-channel exchange of a far off-shell $W$ boson which decays to
produce a top quark and a $b$ antiquark. 

In March 2009, the CDF and D\O\ Collaborations reported
observation of the electroweak production of single 
top quarks in {\ppbar} collisions at $\sqrt{s} = 1.96$~TeV based on
3.2~fb$^{-1}$ (CDF) ~\cite{sitop_obs_cdf} and 2.3~fb$^{-1}$ (D\O)
~\cite{sitop_obs_d0} of 
data. Both 
collaborations used events containing an isolated electron or 
muon and missing transverse energy, together with jets where one or
two of the jets were required to originate 
from the fragmentation of $b$ quarks.

\begin{figure*}[t]
\centering
\includegraphics[width=155mm]{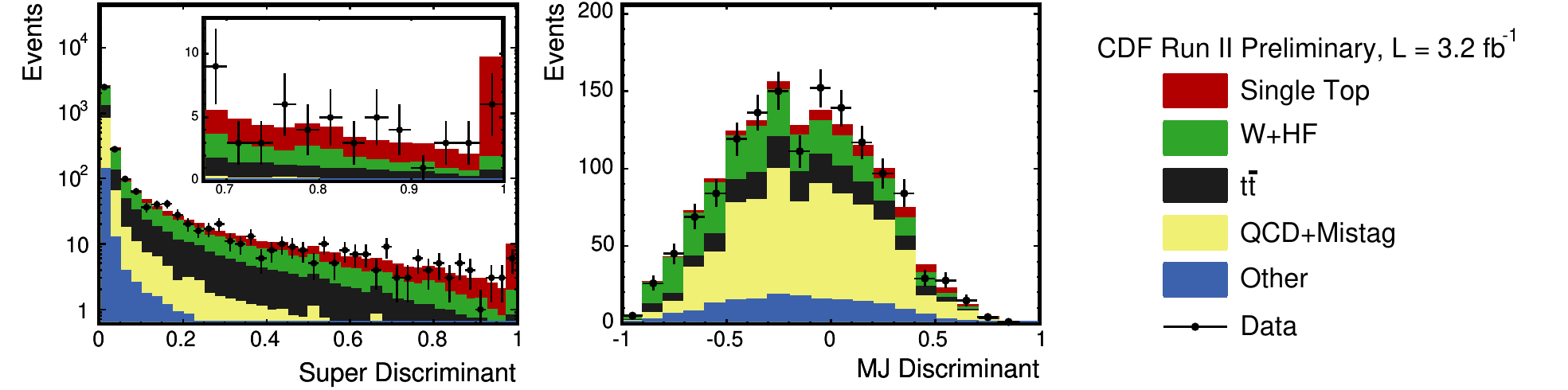}\\
\includegraphics[width=70mm]{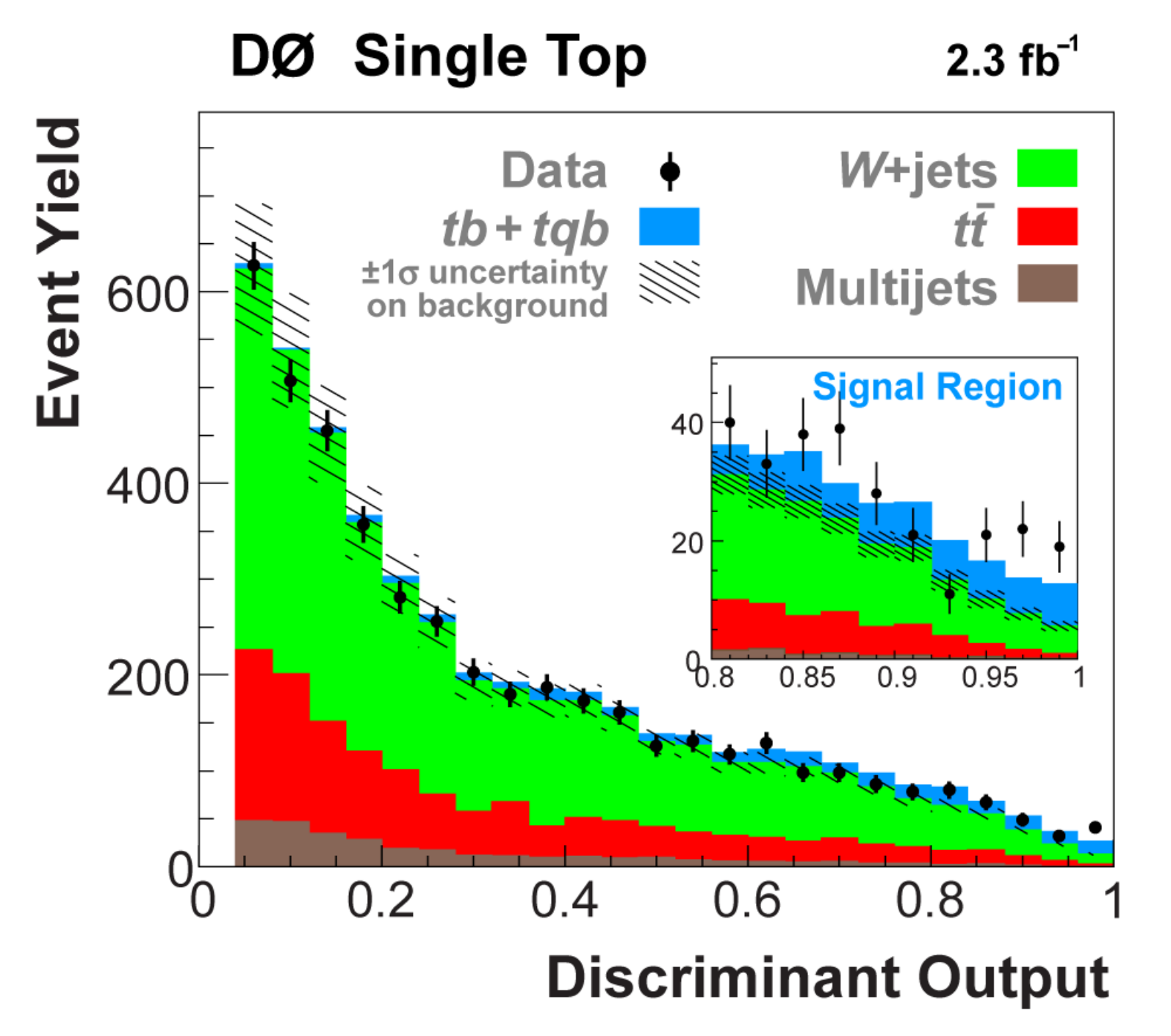}
\includegraphics[width=70mm]{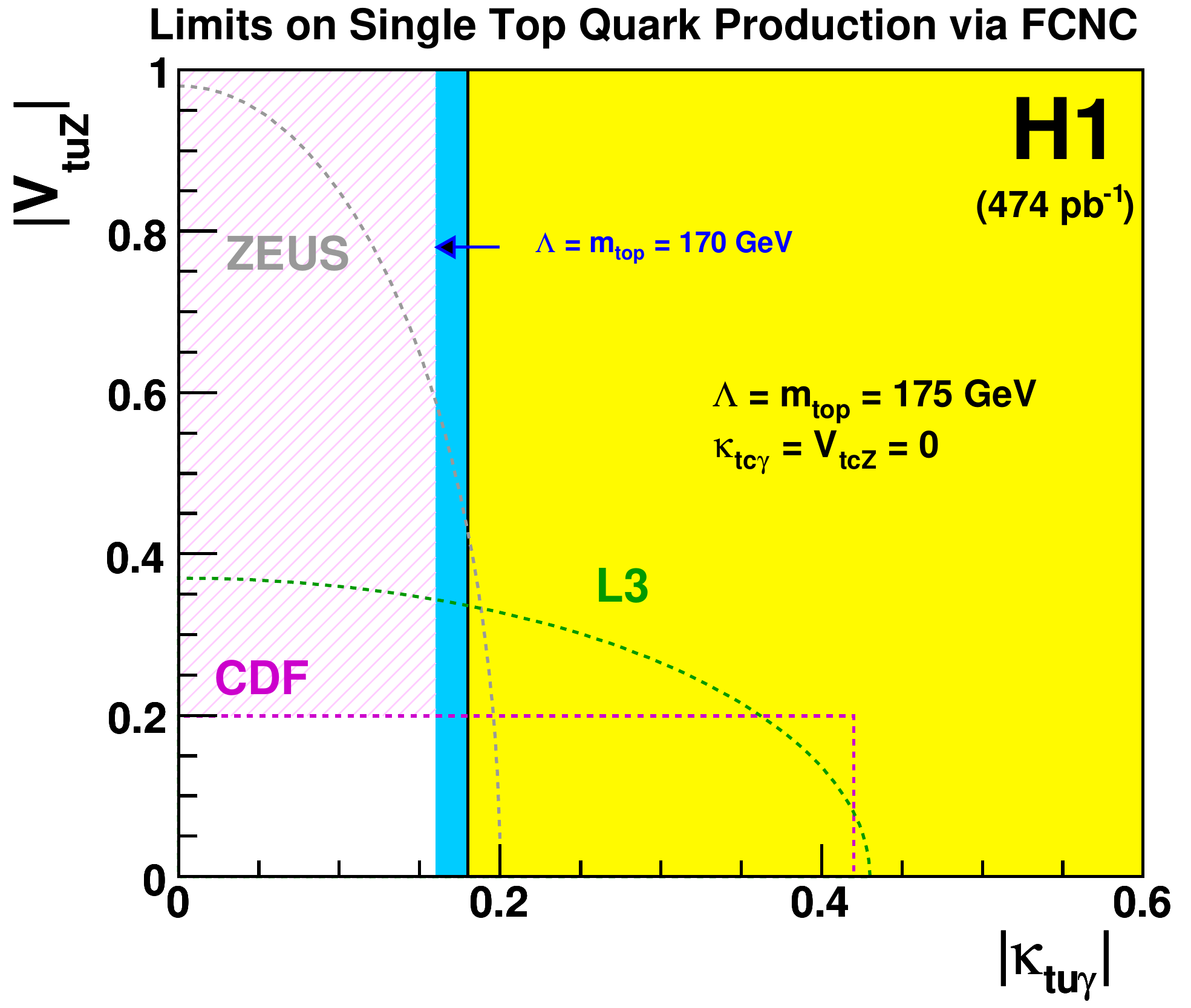}
\caption{Output discriminants for single top quark
  observation. Displayed are the CDF super-discriminant output (upper left),
  the CDF discriminant output for the analysis using a missing transverse
  momentum and jets selection (upper middle) and the D\O\ BNN
discriminant (lower left). 
The lower right plot shows upper limits at 95\% CL on the
anomalous $\kappa_{tu\gamma}$ and $V_{tuZ}$ couplings obtained by the
H1 Collaboration. 
} \label{fig:sitop}
\end{figure*}

CDF and D\O\ each combine many variables using different multivariate
analysis methods such as Boosted Decision Trees, Neural Networks,
Bayesian Neural Networks, Matrix Elements and Likelihood functions to
increase the separation power between signal and background. The
discriminant outputs of each multivariate analysis are combined to one
discriminant taking the correlations into account. CDF combines 8 subchannels into a
super-discriminant using a neural network trained with
neuroevolution~\cite{Nevolution} which is shown in
Fig.~\ref{fig:sitop} (upper left).
The CDF collaboration has an additional independent search
channel~\cite{sitop_obs_cdf,sitop_mj_cdf}
that is designed to select events with missing transverse energy and
jets and is orthogonal to the other channels. It accepts events in
which the $W$ boson decays 
into hadronically decaying $\tau$ leptons and recovers events lost
due to electron or muon identification inefficiencies. The
discriminant output of this analysis -- shown in Fig.~\ref{fig:sitop}
(upper middle) -- is combined to the
super-discriminant to obtain the final result.  D\O\ combines 12
subchannels into one BNN discriminant displayed in
Fig.~\ref{fig:sitop} (lower left).

CDF measures a cross section of
$\sigma({\ppbar} \rightarrow\ tb+X,~tqb+X) =2.3 ^{+0.6}_{-0.5}$~pb assuming
a top quark mass of 175~GeV
and 
D\O\ measures a cross section of
$\sigma({\ppbar} \rightarrow\ tb+X,~tqb+X) = 3.94 \pm 0.88$~pb at a top
quark mass of 170~GeV.
Both measurements correspond to a
5.0~standard deviation (SD) significance for the observation. They are in
agreement with the SM predictions.
Both results are translated into a direct measurement
of the amplitude of the CKM matrix element $V_{tb}$ without making
assumptions on the number of quark generations and the matrix unitarity.
CDF obtains $|V_{tb}| = 0.91 \pm 0.13$, D\O\ derives $|V_{tb}| = 1.07 \pm 0.12$.

The D\O\ collaboration also reported direct evidence for the
electroweak production of single top quarks through the $t$-channel
exchange of a virtual $W$ boson alone~\cite{tchannel_d0}. 
In the observation analyses the combined $s+t$ channel
single top 
quark cross section was measured, assuming the SM ratio of the two production
modes. This ratio is modified in several new physics scenarios,
for example in models with additional quark generations, new heavy
bosons, flavor-changing neutral currents, or anomalous top quark
couplings. Therefore it is interesting to remove this constraint and
to use the $t$-channel 
characteristics to measure the $t$-channel and $s$-channel
cross sections simultaneously which provides a $t$-channel
measurement independent of the $s$-channel cross section model.  

The measured cross section is $\sigma(t-{\rm channel}) =
3.14^{+0.94}_{-0.81}$~pb, has a significance of 4.8 SD and is
consistent with the SM prediction.  
This is the first analysis to isolate an
individual single top quark production channel.

\subsubsection{Single Top Production at HERA}

In $e^\pm p$ collisions at HERA the production of single top quarks is
kinematically possible due to the large center-of-mass energy of up to
$\sqrt{s} = 319$~GeV. Within the SM the production of top quarks at
HERA is however strongly suppressed. Therefore the observation of
single top quark production would be a clear indication of new
physics. In several extensions of the SM the top quark is predicted to
undergo flavor changing neutral current (FCNC) interactions, which
may lead to a sizeable top quark production cross 
section at HERA~\cite{fcnc}. 

A search for single top quark production is performed using the full
$e^\pm p$ data sample collected by the H1 experiment at HERA~\cite{fcnc_h1}. The data
correspond to an integrated luminosity of 474~\pbmone\ including
36~\pbmone\ of data taken at $\sqrt{s} = 301$~GeV. 
Decays of top quarks into a $b$ quark and a $W$ boson with subsequent
leptonic or hadronic decay of the $W$ are investigated. A multivariate
analysis is performed to discriminate top quark production from SM
background processes. An upper limit on the top quark production cross
section via flavor changing neutral current processes $\sigma(ep
\rightarrow etX)< 0.25$~pb is derived at  95\% CL. Limits on the
anomalous couplings $\kappa_{tu\gamma}$ and $V_{tuZ}$ are derived for
$m_t=175$~GeV as depicted in 
Fig.~\ref{fig:sitop} (lower right). The domain excluded by H1 is
represented by the light shaded area. The dark shaded band shows the
region additionally excluded if $m_t$ is varied to 170 GeV. The
hatched area corresponds to the domain excluded at the Tevatron by the
CDF experiment ($\sqrt{s} = 1.96$~TeV, ${\cal L} =
1.9$~\fbmone)~\cite{fcnc_cdf}. Also shown are limits obtained at the 
Large Electron Positron (LEP) Collider by the
L3 experiment ($\sqrt{s} = 189 - 209$~GeV, ${\cal L} =
634$~\pbmone)~\cite{fcnc_l3} and at HERA by the ZEUS
experiment($\sqrt{s} = 300 - 318$~GeV, ${\cal L} =
130.1$~\pbmone)~\cite{fcnc_zeus}. 
This shows nicely the complementarity of the results obtained at
different colliders.

\subsection{Top Quark Properties}

In order to make sure that the discovered top quark is really the top
quark as predicted by the SM its properties have to be measured as
precisely as possible. A few examples of top property measurements
are presented here.

\subsubsection{Top Quark Mass}
The mass of the top quark, which is by far
the heaviest of all quarks, plays an important role in  
electroweak radiative corrections and therefore in constraining the
mass of the Higgs boson. Precise measurements of  
the top quark mass provide a crucial test of the consistency of the
SM and could indicate a hint of  
physics beyond.   

Different methods to measure the top quark mass are discussed.
``Template Methods''~\cite{cdf_allhad} have the
advantage of being more straightforward and transparent but
are statistically less accurate. To maximize the
statistical information on the top quark mass extracted from the event
sample, more elaborated but also more complex methods exist as for
example the 
``matrix method''~\cite{CDFmatrix,D0matrix}. An alternative method uses the cross
section measurement to extract the top quark
mass~\cite{mass_xsec_d0}. Some example analyses are
presented here. 

Distributions of variables that are strongly correlated
with the top quark mass are derived as templates in MC simulations for
different top mass hypotheses. They are compared to the
measured distribution in order to extract the top quark mass from
data. One example is the extraction of the top quark mass with
simultaneous (in situ) calibration of the JES in the all-hadronic
channel using the reconstructed top quark and $W$ boson masses as
templates. The result is $m_t = 174.8 \pm 2.4 {\rm (stat +
  JES)} ^{+1.2}_{-1.0} {\rm (syst)}$~GeV.

However, the currently best measurements use matrix elements to 
calculate a probability for each
event as a function of the assumed top quark mass $m_t$ and an
overall multiplicative scale factor JES for jet energies in
$\ell$+jets final states. The factor
JES is fitted {\em in situ} in data, simultaneously with the top quark
mass by using information from the invariant mass of the hadronically
decaying $W$ boson. For every event, this mass is constrained 
to be equal to the known value for the $W$ mass. The probabilities
from all events in the sample are then combined to obtain a
probability as a function of $m_t$ and JES, and the top quark
mass is extracted by finding the values that maximize this probability. 

\begin{figure*}[t]
\centering
\includegraphics[width=90mm]{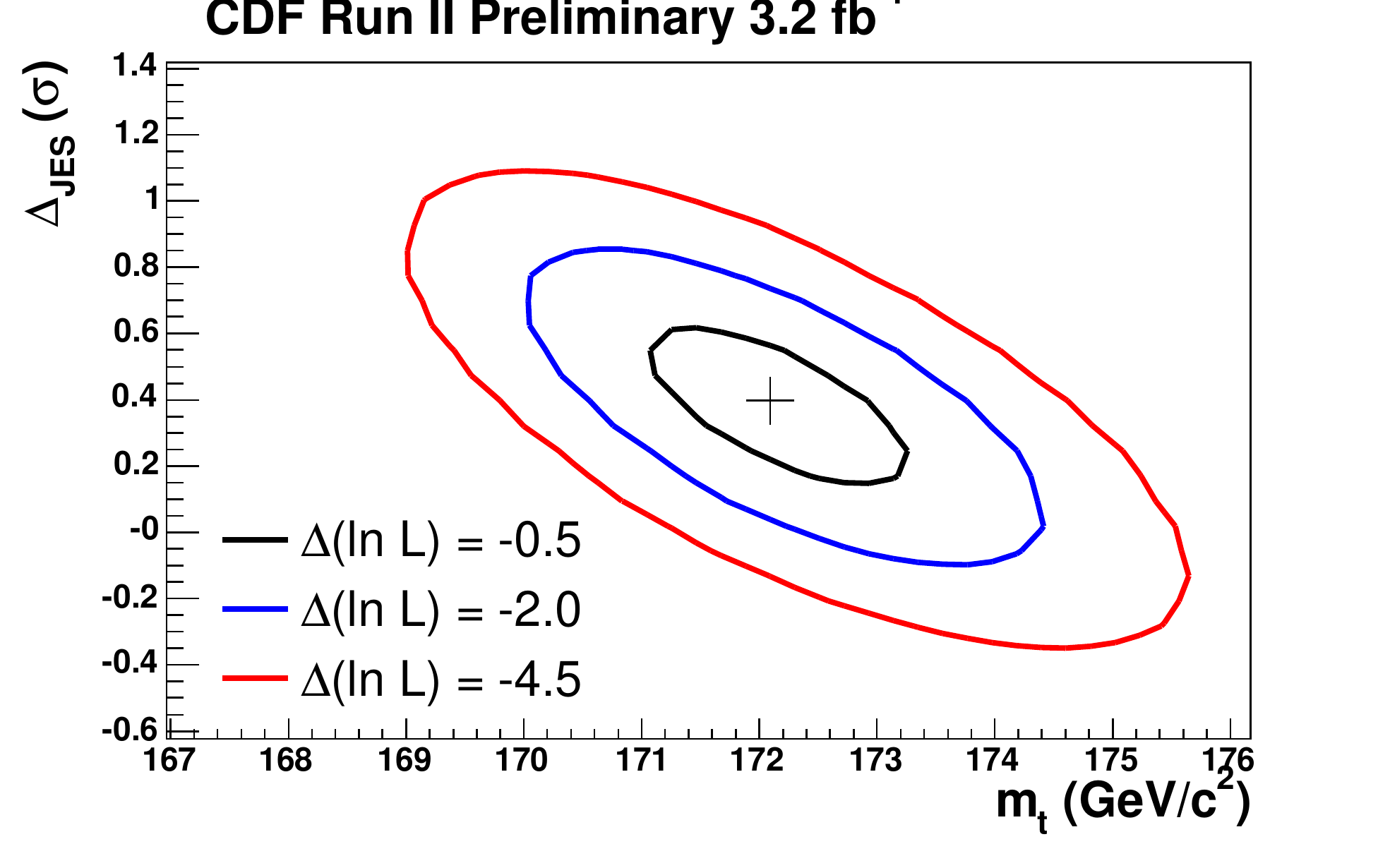}
\includegraphics[width=60mm]{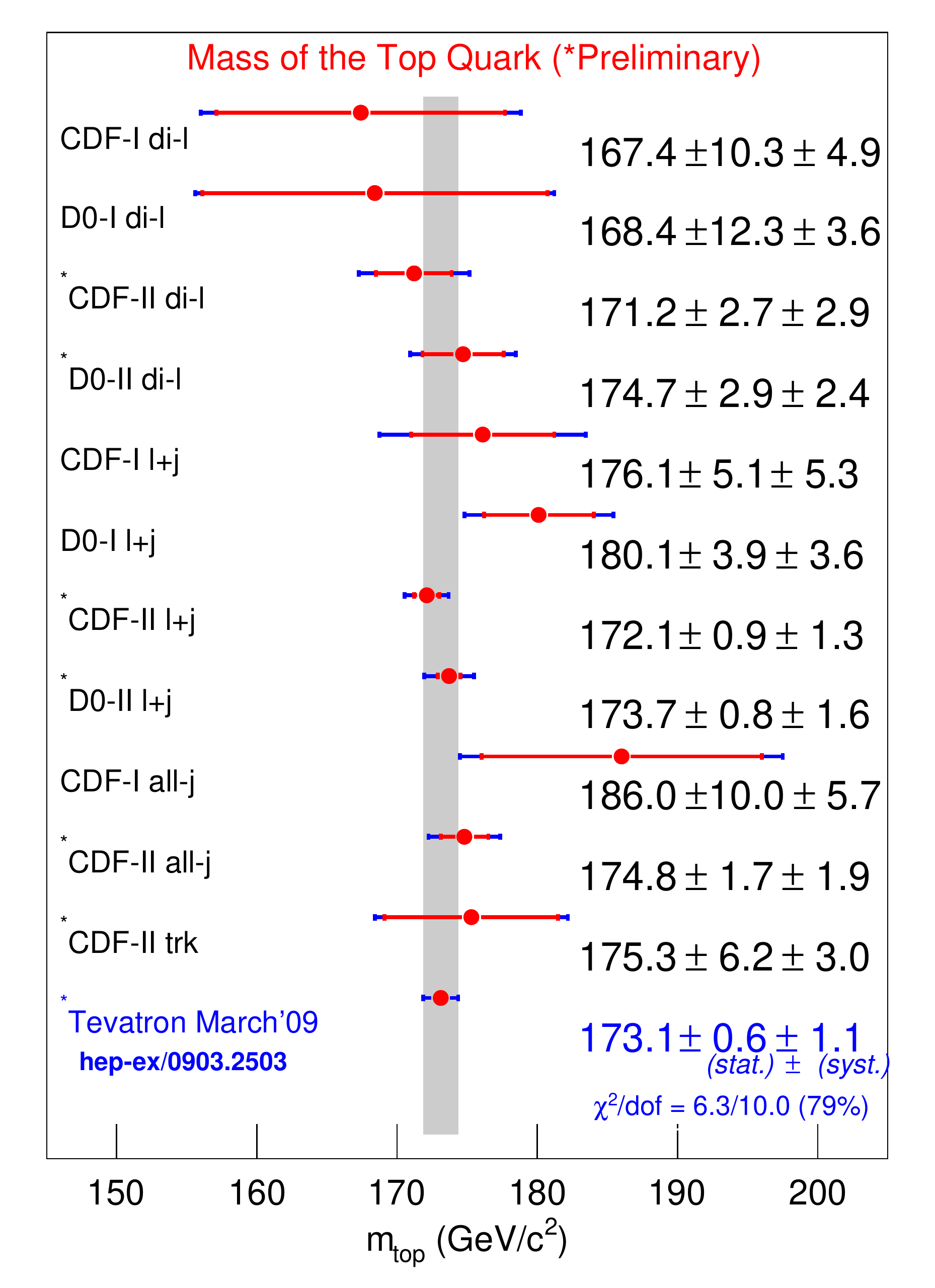}
\caption{2D-likelihood on data events including the $1\sigma$,
  $2\sigma$ and $3\sigma$ uncertainties for the CDF matrix element
  method (left). Summary of all mass measurements
  that are included into the Tevatron combination (right).
} \label{fig:mass}
\end{figure*}

The analyses performed by the CDF~\cite{CDFmatrix} and
D\O~\cite{D0matrix} experiments are very similar.  As a
result the top quark mass is measured to $m_t = 173.7 \pm 0.8
({\rm stat}) \pm 1.6 ({\rm syst})$~GeV by D\O\ and
$m_t = 172.1 \pm 0.9 
({\rm stat}) \pm 1.3 ({\rm syst})$~GeV by CDF.
For the latter measurement, the contours of the log likelihood
function for the data are shown as a 
function of the top quark mass and JES in Fig.~\ref{fig:mass} (left).
The
total uncertainties are $\pm1.0$\% (D\O) and $\pm0.9$\% (CDF), respectively. They are
systematically limited. The largest sources of systematic
uncertainties apart from the simultaneous inclusion of JES are given
by residual JES, in particular of $b$-jets, and theoretical
uncertainties in signal and background modeling. Currently both
experiments are undertaking large efforts to get a uniform treatment
of all uncertainties where ever possible.

Taking correlated uncertainties properly into account the CDF and
D\O\ collaborations have derived a combination 
of published Run-I (1992--1996) measurements with the most recent 
preliminary Run-II (2001--present) measurements using up to $3.6\ {\rm
  fb}^{-1}$ of data as represented in Fig.~\ref{fig:mass} (right). A new world
average~\cite{worldave} of $m_{\rm 
  top} = 173.1 \pm 0.6   
({\rm stat}) \pm 1.1 ({\rm syst})$~GeV was obtained. This corresponds
to a total precision of 1.3~GeV and a relative
precision of 0.75\% on the top quark mass.

\subsubsection{Top Antitop Quark Mass Difference}
The	CPT theorem, which is fundamental to any local
Lorentz-invariant quantum field theory, requires that 
the mass of a particle and that of its antiparticle be identical.
The D\O\ collaboration reported a measurement of the difference
between the mass of 
the top quark and that of its antiparticle based on data corresponding
to 1~\fbmone\ of integrated luminosity~\cite{massdiff_d0}.

\begin{figure}[h]
\label{fig:mean-template}
\begin{centering}
\includegraphics[scale=0.32]{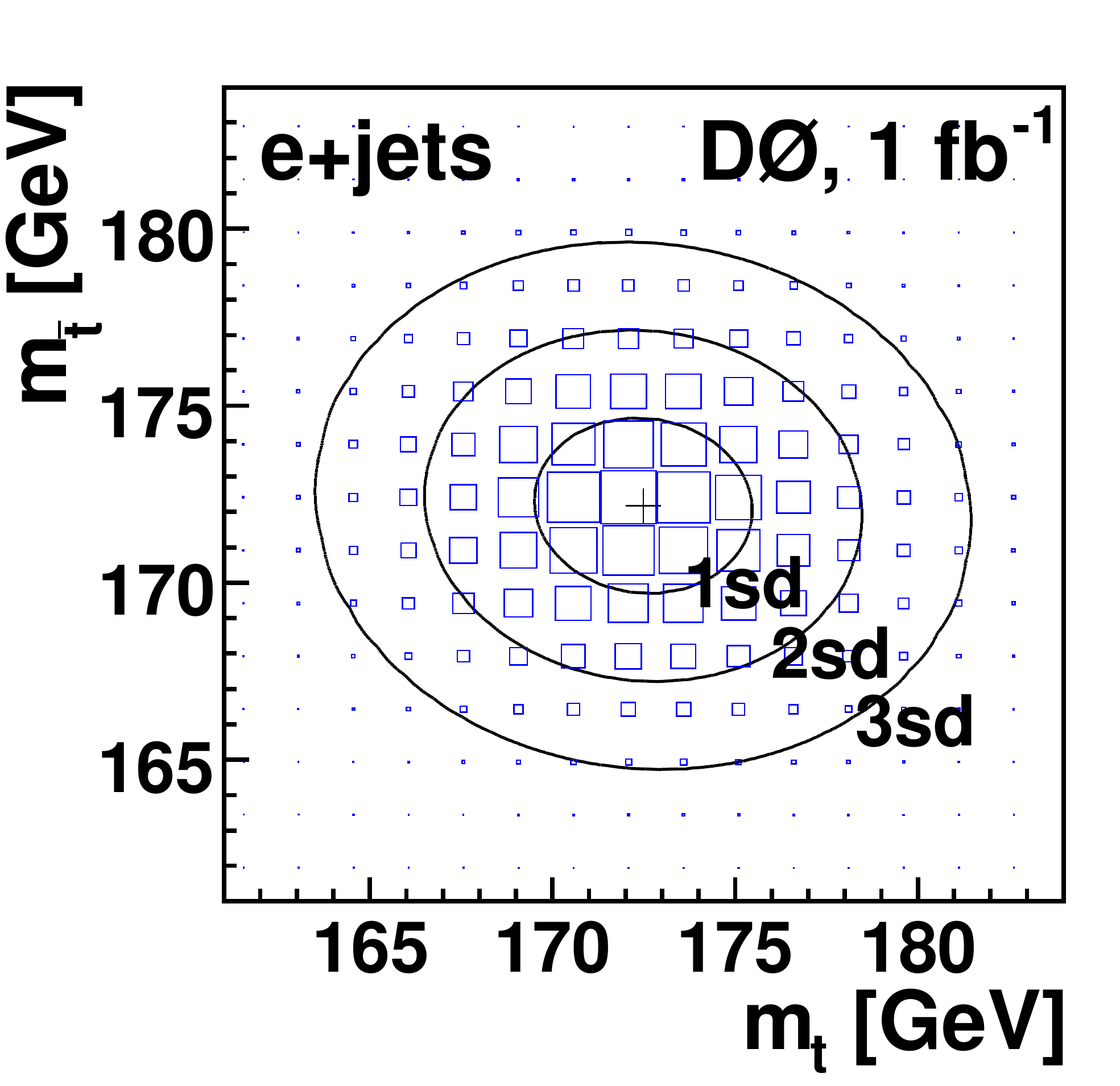}
\includegraphics[scale=0.43]{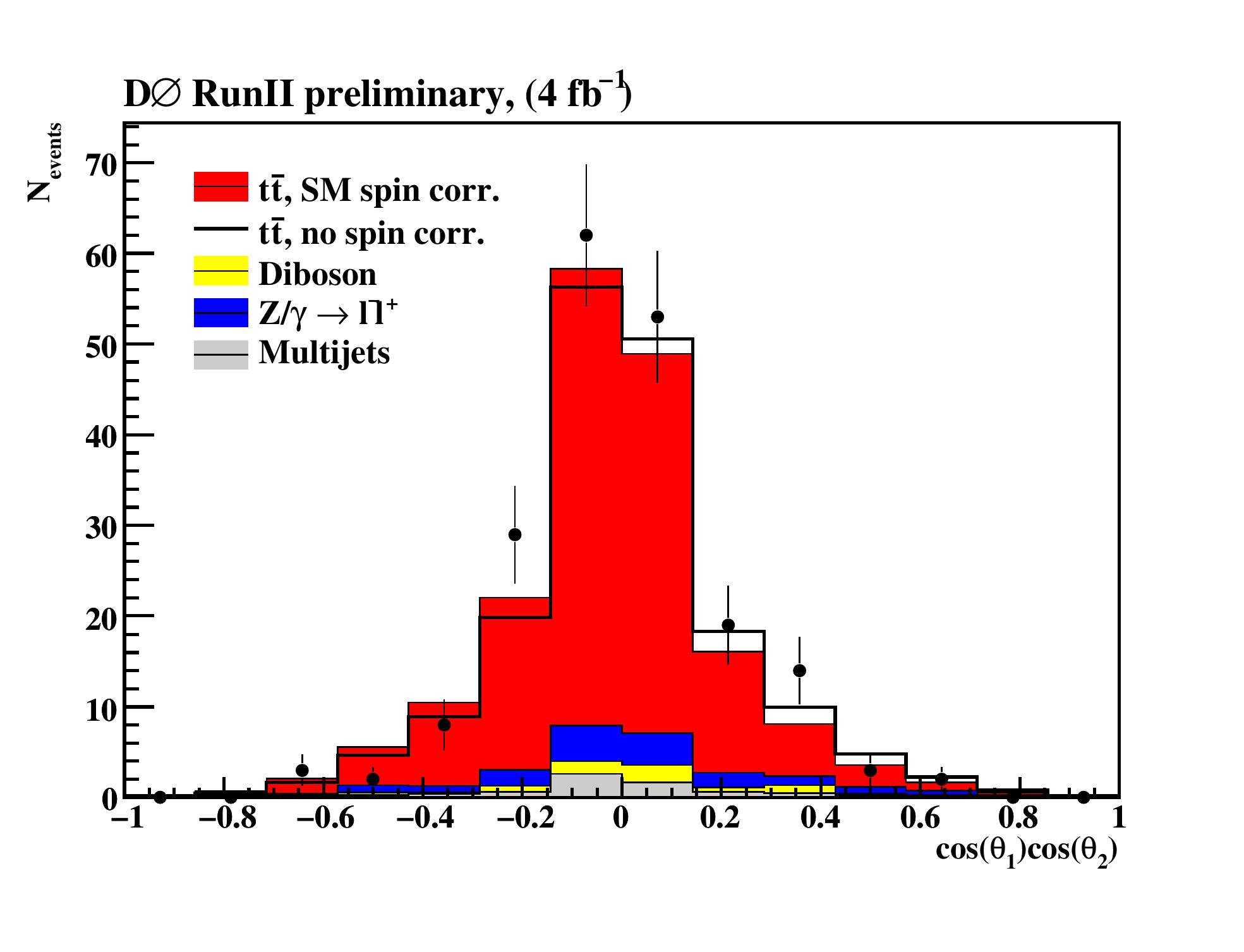}
\par\end{centering}

\caption{\label{fig:massdiff_spin}
Left: fitted contours of equal probability for the two-dimensional
likelihoods as a function of $m_t$ and $m_{\bar{t}}$ in $e$+jets
data. The boxes, representing the bins in the two-dimensional
histograms of the likelihoods, have areas proportional to the bin
contents, set equal to the value of the likelihood evaluated at the
bin center. 
Right:
the $\cos\left(\theta_{1}\right)\cos\left(\theta_{2}\right)$
distribution for a dilepton event sample. The sum of \ttbar\ signal including NLO QCD spin correlation
($C=0.78$) (red) and multijet (grey), di-boson (yellow) and Drell-Yan
(blue) background is compared to data. The open black histogram shows
the prediction without \ttb\ spin correlation ($C=0$).}

\end{figure}

This analysis uses a similar matrix element method as the analyses described
in the previous section. Here the jet energies are
scaled by an overall JES calibration factor and the mass of the top quark and the mass of
the antitop quark, which are not constrained to be equal anymore, are
extracted simultaneously. This is
shown in Fig.~\ref{fig:massdiff_spin} (left) for the $e$+jets channel. The
mass difference for the combined $\ell$+jets channels was obtained to
be $m_t - m_{\bar{t}} = 3.8 \pm 3.7$~GeV. This is the first direct measurement of a mass difference
between a bare quark and its antiquark partner.

%
%

\subsubsection{Top Antitop Quark Spin Correlation}
\label{sec:spin}

One of the most important
properties of the top quark, its spin, is still basically
unexplored.
While the top quarks and antiquarks produced at hadron collider are
unpolarized, their spins are correlated~\cite{Barger:1988jj,Stelzer:1995gc}.
Since at the Tevatron top pair production is dominated by $q\bar{q}$
scattering, a different spin correlation is analyzed compared to the LHC
where top pair production is dominated by $gg$ scattering.

The SM predicts that the top quark decays before its spin
flips~\cite{Bigi:1986jk},
in contrast to the lighter quarks, which are depolarized by QCD
interactions long before
they fragment~\cite{Falk:1993rf}. The spin of the top quark is therefore
reflected by its decay products. In the following analyses it is assumed that
top quarks decay exactly as predicted by the SM. Then the charged lepton from
a leptonic top quark decay has a spin analyzing power of 1 at the
tree level~\cite{Brandenburg:2002xr}. Therefore, the dilepton final
states have the highest
sensitivity to measure the correlation between the spins of
pair-produced top and antitop quarks.

The only measurement of spin correlations between top and
antitop quarks so far was performed by the \dzero\
Collaboration using an integrated luminosity of approximately
125~\pbmone\ of \ppbar\ collisions at $\sqrt{s}=1.8$~TeV during
Run-I of the Tevatron collider~\cite{D0RunI}. The 68\% confidence level
given on a correlation coefficient was in agreement with the SM prediction.
However, since the sample used for the measurement contained only six
events the sensitivity was too low to distinguish between a hypothesis of no
correlation between the spins of top and antitop quarks and the
correlation predicted by the SM.

The CDF and D\O\ collaborations have measured the spin correlations
between $t$ and $\bar{t}$ 
now for the first time in proton antiproton
scattering at a center of mass energy of $\sqrt{s}=1.96$~TeV using
data of an integrated luminosity of 2.8~\fbmone\ (CDF) and up to 4.2~\fbmone\ (D\O), respectively.

The following double differential
distribution is used:
%
\begin{equation}
\label{eq:coscos}
\frac{1}{\sigma} \frac{d\sigma}{d\cos\theta_1 d\cos\theta_2} =
\frac{1}{4} ( 1 - C \cos\theta_1 \cos\theta_2 ) \,\, ,
\end{equation}
%
where $\sigma$ denotes the cross section of the channel under
consideration and $C$ is a free parameter between -1 and 1 that
depends on the choice 
of the spin basis. 
At tree level in the SM $C$ represents the number of events where the $t$ and
$\bar{t}$ spins are parallel minus the number of events where they are
anti-parallel normalized by the total number of events.
$\theta_1$ ($\theta_2$) describes the angle between 
the direction of flight of the lepton $\ell^+$ ($\ell'^-$) in the $t$
($\bar{t}$) rest frame and a reference direction $\bf \hat{a}$ ($\bf
\hat{b}$). The choice of the spin basis determines the extent to which the
spins of the top quarks are correlated. For the Tevatron it has been
shown that an almost optimal choice for the spin 
basis is given by the beam basis~\cite{spin_theory}.

Templates are generated for different values of $C$ 
by reweighting generated MC events at parton level.
D\O\ uses the distribution (\ref{eq:coscos}) which
is shown in Fig.~\ref{fig:massdiff_spin} (right)
for background and \ttbar\
signal without and with NLO QCD spin correlation. 
CDF uses the
double-differential $(\cos\theta_1, \cos\theta_2 )$ and
$(\cos\theta_b, \cos\theta_{\bar{b}} )$ distributions where for the
latter the flight directions of $\ell^+$ ($\ell'^-$) are replaced by
those of the $b$ quark (anti-$b$ quark). D\O\ uses the beam axis as
the quantization axis and CDF the off-diagonal basis.

The extracted result is
$C_{\rm beam} = -0.17^{+0.64}_{-0.53} \, ({\rm stat + syst})$ in case of the D\O\
collaboration and $C_{\rm off-diagonal} = 0.32^{+0.55}_{-0.78} \,
({\rm stat + syst})$  for CDF. 
This agrees with the NLO QCD value of $C = C_{\rm beam} = C_{\rm
  off-diagonal} =0.78$ within 2 SD.

\subsection{Searches for New Physics in Top Production and Decay}

The fact that the top quark has by far the largest mass of all known elementary
particles suggests that the it may play a special role in
the dynamics of electroweak symmetry breaking. This makes searches for
new physics in the top quark sector very attractive. In the following
a few examples are presented for searches for new physics in the
production and decay of top quarks.

\subsubsection{Search for \ttbar\ Resonances}

One of the various models incorporating the possibility of a special
role of the top quark in the dynamics of electroweak symmetry breaking
is topcolor~\cite{topcolor}, where the large top quark mass can be generated
through a dynamical \ttbar\ condensate, $X$, which is formed by a new strong
gauge force preferentially coupled to the third generation of
fermions. In one particular model, topcolor-assisted technicolor~\cite{Zprime},
$X$ couples weakly and symmetrically to the first and second generations
and strongly to the third generation of quarks, and has no couplings
to leptons, resulting in a predicted cross section for \ttbar\ production
larger than SM prediction. 

The CDF and D\O\ collaborations presented updated
model-independent searches for a narrow-width heavy resonance $X$
decaying into \ttbar\ using 
2.8~\fbmone\ of data in
the all-hadronic final 
state (CDF)~\cite{resonances_cdf}
and 
3.6~\fbmone\ of data analyzing $\ell$+jets
final states (D\O)~\cite{resonances_d0}.
The search for resonant production is
performed by examining 
the reconstructed \ttbar\ invariant mass distribution resulting from a
constrained kinematic fit to the \ttbar\ hypothesis. No deviation from
the SM prediction was observed.
Within a top-color-assisted technicolor model, the existence of a
leptophobic $Z'$ boson with $M_{Z'} < 805$~GeV (CDF) and $M_{Z'} <
820$~GeV (D\O), respectively, and width around $\Gamma_{Z'}	=
0.012 M_{Z'}$ is excluded at 95\% CL.

\subsubsection{Forward Backward Asymmetry}

In LO QCD, the top quark production angle is symmetric with respect to the
beam direction. At NLO, QCD predicts a small charge asymmetry, $A_{fb}
= 0.050 \pm 0.015$~\cite{asym_theo1, asym_theo2}, due to interference
of initial-state radiation diagrams with final-state diagrams and ``box
diagrams'' with Born processes.

\begin{figure*}[t]
\centering
\includegraphics[width=70mm]{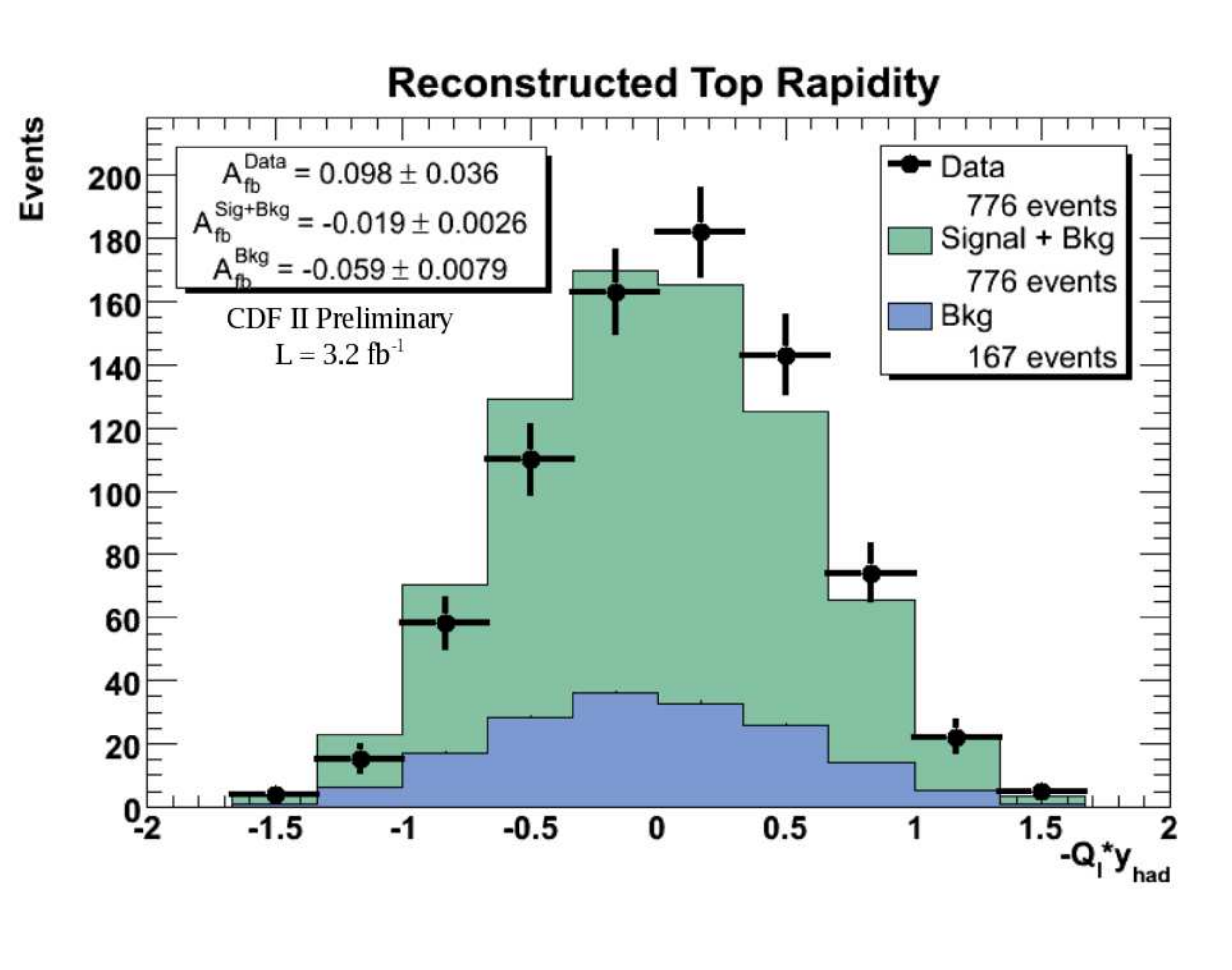}
\includegraphics[width=80mm]{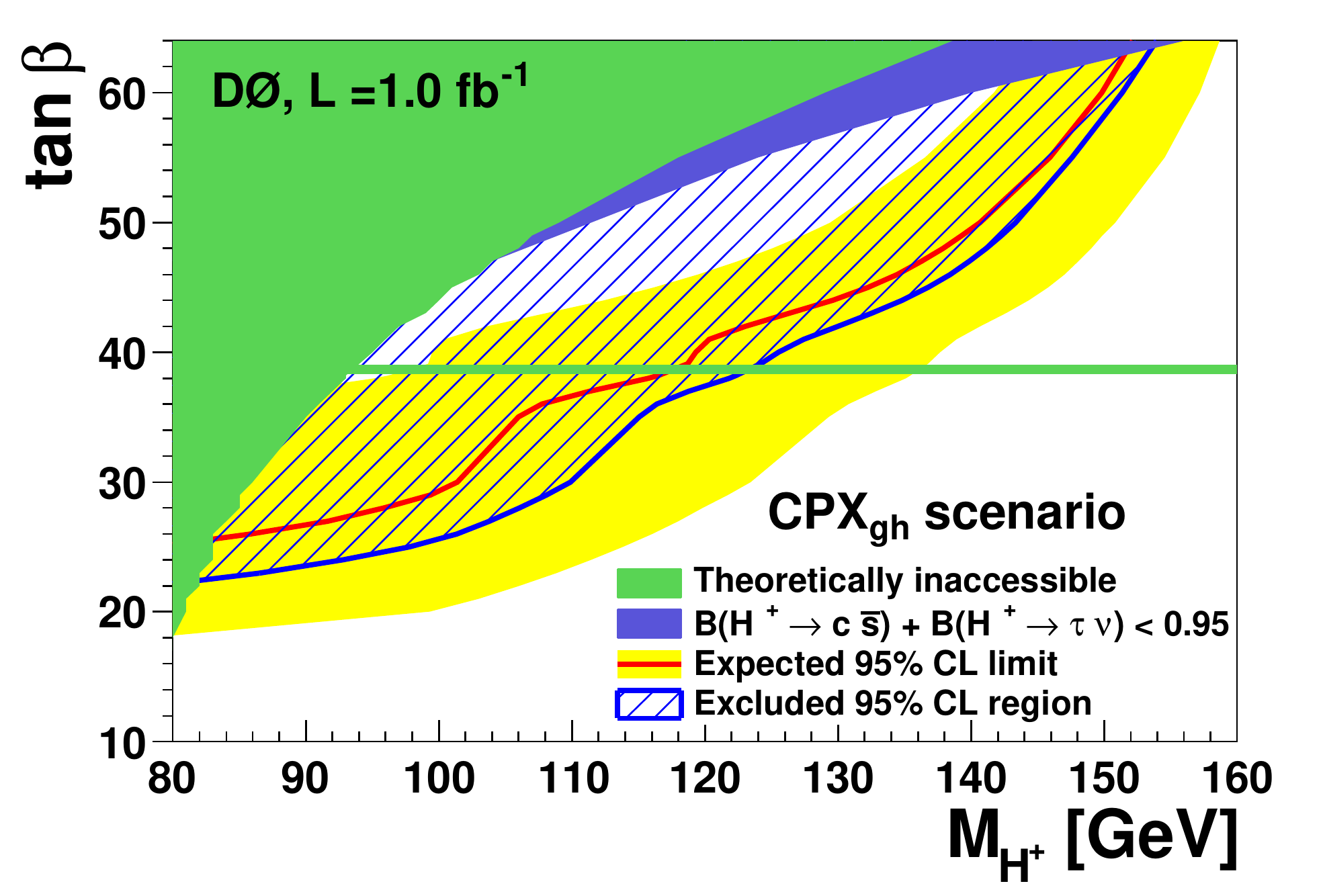}
\caption{Left: the raw distribution $-Q_l \cdot y_{\rm
  had}$ used for the forward-backward asymmetry measurement.
Right: excluded regions in $\tan\beta$ and $M_{H^{+}}$ in a CPX
benchmark scenario of a strangephilic Higgs sector.
} \label{fig:asym_hplus}
\end{figure*}

The CDF collaboration has measured the forward-backward asymmetry of
pair produced top quarks in a dataset of 3.2~\fbmone\ in $\ell$+jets
final states~\cite{asym_cdf}. The \ttbar\ events were reconstructed with a $\chi^2$
based kinematic fitter. Then, in the \ppbar\ laboratory frame, the
rapidity $y_{\rm had}$ of the 
hadronically-decaying top (or antitop) system was studied, tagging the
charge with the lepton sign $Q_l$ from the leptonically decaying
system. The raw distribution of $-Q_l \cdot y_{\rm
  had}$ is shown in Fig~\ref{fig:asym_hplus}
(left). Assuming CP-invariance the asymmetry
\begin{equation}
A_{fb} = \frac{N(-Q_l \cdot y_{\rm had} >0) - N(-Q_l \cdot y_{\rm had} <0)}
{N(-Q_l \cdot y_{\rm had} >0)+N(-Q_l \cdot y_{\rm had} <0)}
\end{equation}
is measured. A model-independent correction for acceptance and
reconstruction dilutions is performed to be able to compare the asymmetry
directly with theoretical predictions. The result is
$A_{fb} = 0.193 \pm 0.065 ({\rm stat}) \pm 0.024  ({\rm syst})$
consistent with the prediction in NLO QCD. The D\O\ asymmetry
measurement is also in agreement with the SM prediction~\cite{asym_d0}. 

\subsubsection{Search for Charged Higgs Bosons in Top Quark Decays}

In many extensions of the SM,
including supersymmetry and grand unified theories, the existence of 
an additional Higgs doublet is required. Such models predict multiple physical 
Higgs particles, including three neutral and two charged Higgs bosons
($H^\pm$) \cite{theory_review}. If a charged Higgs boson is sufficiently light, 
it can appear in top quark decays $t\rightarrow
H^{+}b$~\footnote{Throughout the paper, $H^{+}$ also refers
  to the charge conjugate.}. 
The decay modes of the charged Higgs boson depend on the
ratio of the vacuum expectation values of the two Higgs doublets,
$\tan\beta$. For small values of $\tan\beta$ it is dominated by the decay 
to quarks, while for larger values of $\tan\beta$ 
it is dominated by
the decay to a $\tau$ lepton and a neutrino. 

The D\O\ Collaboration presented a search for charged Higgs bosons
in top quark decays~\cite{hplus_d0}. The \eplus, \muplus, $ee$, $e\mu$,  
$\mu\mu$, \etau\, and \mutau\, final states from top quark pair production
events were analyzed, using data from about 1~\fbmone\ of integrated
luminosity.
Different scenarios of possible charged Higgs boson
decays were considered, one where the charged Higgs boson decays purely hadronically
into a charm and a strange quark, another where it decays into a $\tau$ lepton and a
$\tau$ neutrino and a third one where both decays
appear. 
For a non-zero branching ratio
$B(t\rightarrow H^{+}b \rightarrow c\bar{s} b)$
the number of events decreases in the
\ljets, $\ell\ell$ and \ltau\ final states. 
In case of a non-zero branching ratio
$B(t\rightarrow H^{+}b \rightarrow \tau^+\nu b)$ the number of predicted
events increases in the \ltau\ channel while it decreases in all other
channels. 

No
indication for charged Higgs boson production in the tauonic or leptophobic 
model is found. Comparing the number of predicted and observed events 
in the various $t\bar{t}$ final states, limits on the branching ratio $B(t\rightarrow H^+
b)$ were extracted for all these models. Two methods were used, one 
where the $t\bar{t}$ 
production cross section is fixed, and one where the cross section is
fitted simultaneously with $B(t\rightarrow H^+b)$.  
Based on the extracted limits, regions in the
charged Higgs boson 
mass and $\tan\beta$ were excluded for different scenarios of the minimal
supersymmetric standard model.

As an example a specific CP-violating benchmark scenario (CPX)  
with large threshold corrections~\cite{jae_pilaf} is presented where
values of 
$B(H^{+} \rightarrow \tau^{+} \nu) +B(H^{+}
\rightarrow c \bar{s}) \approx 1$ can be realized introducing an
additional mass hierarchy  
between the first two and the third generation of sfermions.
Figure~\ref{fig:asym_hplus} (right) shows the excluded 
region in the $[\tan\beta,M_{H^{+}}]$ parameter space. Theoretically 
inaccessible regions indicate parts of the parameter space where 
perturbative calculations can not be performed reliably.  
In the $[\tan\beta,M_{H^+}]$ region 
analyzed here, the sum of the branching ratios was found to be
$B(H^{+} \rightarrow \tau^{+} \nu) +B(H^{+} 
\rightarrow c \bar{s})>0.99$ except for values very close to the blue
region which indicates $B(H^{+} \rightarrow \tau^{+} \nu) +B(H^{+}
\rightarrow c \bar{s})<0.95$. The charged Higgs decay 
$H^{+} \rightarrow \tau^{+} \nu$ dominates for $\tan\beta$ below 22 
and above 55. For the rest of the $[\tan\beta,M_{H^{+}}]$ parameter 
space both the hadronic and the tauonic decays of
charged Higgs bosons are important. In the region $38 \le \tan\beta \le 40$,
the hadronic decays of the charged Higgs boson dominate and 
$ B(H^{+} \rightarrow c \bar{s}) > 0.95$. Here the Higgs sector is
strangephilic. The neutral Higgs boson would preferably decay
into an \ssbar\ pair and therefore be missed in
searches using either $b$-tagging or tau identification. In this case
only strangephilic charged Higgs bosons from top decays as analyzed here could be
easily discovered. 
For large values of $\tan\beta$, $M_{H^{+}}$ up to 154~GeV 
are excluded. For low charged Higgs masses, $\tan\beta$ values down to 23 are 
excluded. These are the first Tevatron limits on
a CP-violating MSSM scenario derived from the charged Higgs 
sector.

\section{Prospects at the LHC}

At the time of this conference it was clear that the Large Hadron
Collider (LHC) at CERN would soon
start the first proton-proton collisions at a center of mass energy
never reached before. The investigation of the physics of the top quark will be a
fundamental element of the early analysis program. 
A few highlights of the expectations of top
quark measurements at the ATLAS and CMS 
experiments are given in the
following.

It is expected that at a luminosity of $10^{32} {\rm cm}^{-2} {\rm
  s}^{-1}$ and a center-of-mass energy of 10~TeV, a few weeks will be
sufficient to rediscover the top quark and to collect a sample large
enough to check SM predictions, but also to test the validity of
lepton identification, jet identification, $b$-tagging and many other
analysis tools in general. 

\begin{figure*}[t]
\centering
\includegraphics[width=75mm]{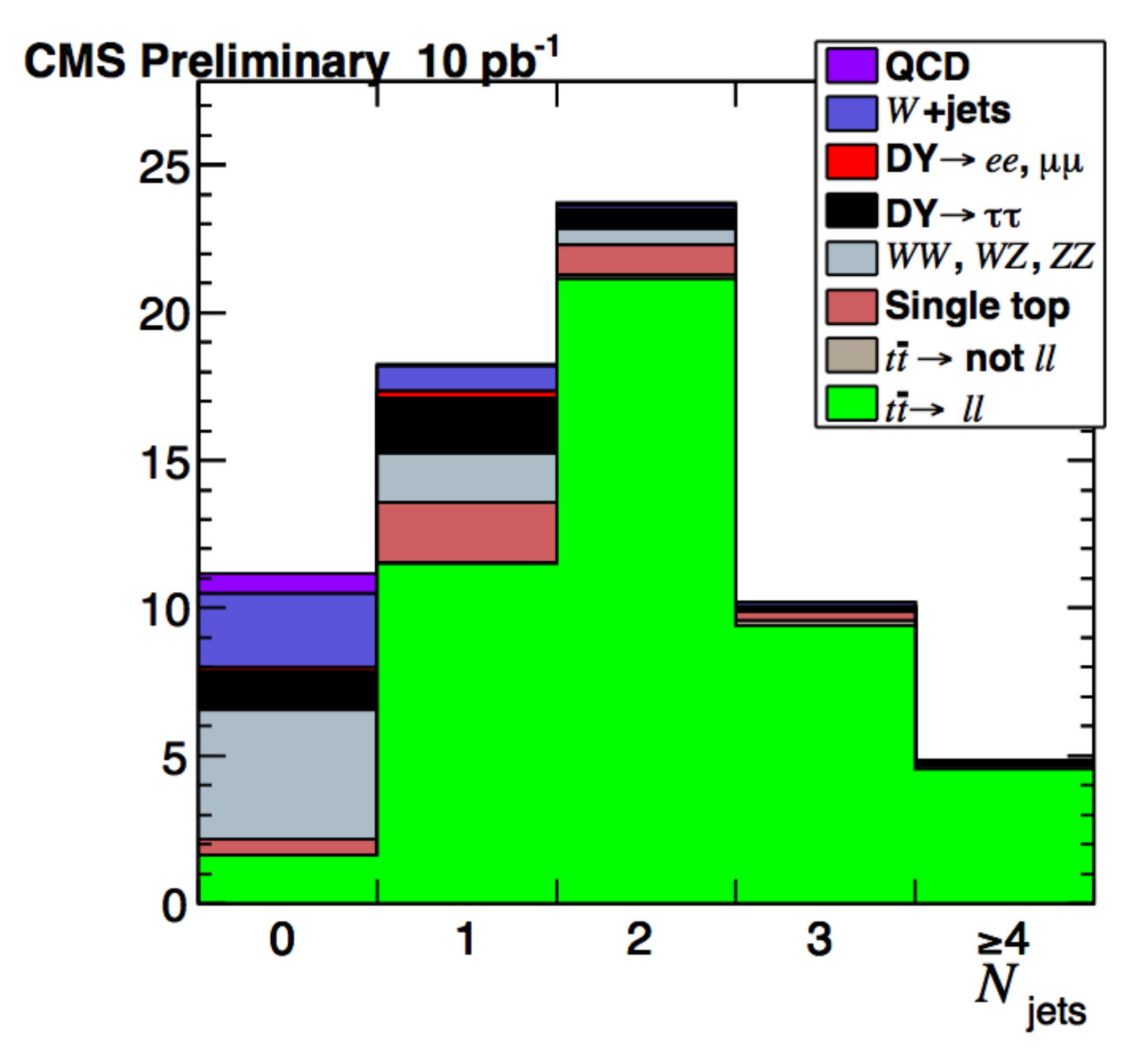}
\includegraphics[width=75mm]{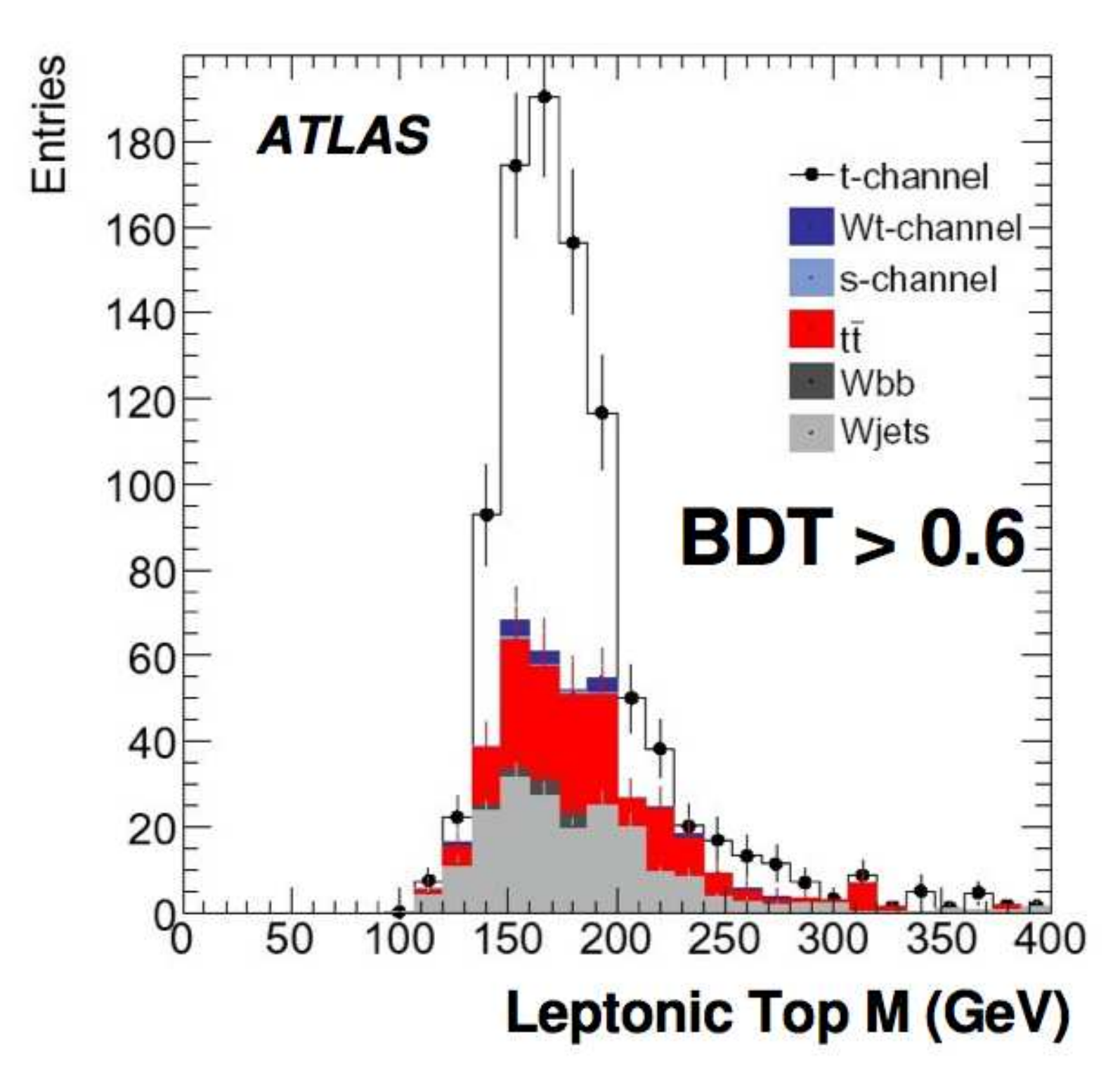}
\caption{Left: number of jets in a dilepton selection with a dataset
  of 10~\pbmone\ at $\sqrt{s}=10$~TeV as expected by CMS. Right:
  reconstructed mass of 
  leptonically decaying single top quarks produced by $t$-channel
  $W$ exchange
  with a dataset of 1~\fbmone\ at $\sqrt{s}=14$~TeV as expected by ATLAS.
} \label{fig:xsec_lhc}
\end{figure*}

As a first step the top pair production will be
measured~\cite{tt_atlas_prospect,tt_cms_prospect}. As an example, the
CMS collaboration 
expects to measure the \ttbar\ cross section with an uncertainty of
$\Delta\sigma_{\ttbar} = 15\% {\rm (stat)}  \pm 10\% {\rm (syst)}  \pm
10\% {\rm (lumi)}$ with a dataset of 10~\pbmone\ at
$\sqrt{s}=10$~TeV~\cite{tt_cms_prospect}. 
Figure~\ref{fig:xsec_lhc} shows the expected distribution of the number
of jets for the \ttbar\ signal and the different sources of
background.

ATLAS and CMS also evaluated the prospects for the measurement of the
single top cross section and the direct extraction of
$V_{tb}$~\cite{sitop_atlas_prospect,sitop_cms_prospect}. As an example, 
ATLAS studied the measurement of $V_{tb}$ in the $t$-channel
based on a multivariate Boosted Decision Tree (BDT)
technique~\cite{sitop_atlas_prospect}. Figure~\ref{fig:xsec_lhc} shows
the expected reconstructed 
mass of the leptonically decaying top quark for large values of the
BDT discriminant where the signal is enhanced. Assuming a center-of-mass
energy of 14~TeV and an  
integrated luminosity of 1~\fbmone, the
expected accuracy is $\Delta
V_{tb} = 12\% {\rm (stat+syst+theo)}$ which is dominated by systematic
uncertainties due to $b$-tagging, JES and
luminosity. 
This compares to the current
measurements of $\Delta V_{tb} = 14\%$ (CDF) and $\Delta V_{tb} =
11\%$ (D\O) at the Tevatron.

In general, it will be difficult for the LHC
experiments to beat systematically limited Tevatron measurements of properties such as
the top quark mass. Here the expectation of 
an accuracy of 1~GeV at the LHC is very close to the current Tevatron
accuracy of 1.3 GeV which will further improve with more data and an
even better understanding of the detectors. However, the LHC will be
able to measure basic quantities
such as spin, charge and the couplings of the top quark with a very
high precision in the future. Most interestingly, some quantities will
be measured for the first time. One example is the spin
correlation between top and antitop quarks where the LHC
measures a different quantity compared to the Tevatron, because the
top pairs are mostly 
produced by gluon fusion while \qqbar\ annihilation dominates at the
Tevatron. The ATLAS collaboration has presented
a study for the $\ell$+jets channel which shows an expected accuracy
of 53\% for a dataset of only 220~\pbmone\ at $\sqrt{s}
=14$~TeV~\cite{sitop_atlas_prospect}. 

In many searches for new physics the LHC experiments have a much better
expected sensitivity even with early data compared
to the Tevatron analyses. One example is the search for
FCNC where the expected limit in absence of a signal will exceed all 
LEP, HERA and Tevatron limits shown in Fig.~\ref{fig:sitop} (lower
right) by several factors 
using ATLAS data corresponding to an integrated luminosity of only
1~\fbmone\ at $\sqrt{s} =14$~TeV~\cite{sitop_atlas_prospect}.  

Furthermore, exciting new developments were presented such as a method for the
discrimination of boosted hadronically decaying top jets against light
quark and gluon jets using jet substructure. The
procedure presented by the CMS collaboration involves parsing the jet
cluster to resolve its subjets, and then imposing kinematic
constraints due to for example the top quark or $W$ boson masses. 
With this method, light quark or gluon jets with a transverse momentum
of around 1 TeV can be rejected with an efficiency of around 99\%
while retaining up to 40\% of top jets~\cite{Kaplan:2008ie}.

\section{Conclusions}

Recent highlights in top quark physics analyses from the HERA and Tevatron
colliders and prospects from the LHC are reviewed. Among many impressive results
from the Tevatron, the 
observation of single top quark production via the electroweak interaction
and the direct measurement of the CKM matrix element $V_{tb}$ by the
CDF and D\O\ Collaborations is
outstanding. 

Analyzing datasets corresponding to an integrated
luminosity of up to 3.6~\fbmone\ the Tevatron experiments can now
perform several high precision measurements, most of all of the top quark
mass which is known to an accuracy of 0.75\% in its current world
average. For the first time, sensitive measurements of important and
previously unexplored 
properties such as \ttbar\ spin 
correlations could be performed.
In all presented analyses, so far, there is no hint for new physics beyond the SM in the
top sector. Excellent prospects for top physics investigations can be
expected with more integrated luminosity at the Tevatron and by the
upcoming LHC physics run.

\end{document}